\documentclass[%
 aip,
 amsmath,amssymb,
 reprint,%
]{revtex4-1}

\usepackage{graphicx}
\usepackage{dcolumn}
\usepackage{bm}
\usepackage{balance}
\usepackage{mhchem}
\usepackage[LGR,T1]{fontenc}
\usepackage{textcomp}
\usepackage{tipa}
\usepackage{amstext}
\usepackage{subscript}
\usepackage{times,mathptmx}
\usepackage{lastpage}
\usepackage{float}
\usepackage{fancyhdr}
\usepackage{fnpos}
\usepackage[english]{babel}
\usepackage{array}
\usepackage{droidsans}
\usepackage{charter}
\usepackage[T1]{fontenc}
\usepackage[usenames,dvipsnames]{xcolor}
\usepackage{setspace}
\usepackage[compact]{titlesec}
\usepackage[colorlinks,
            linkcolor=black,
            anchorcolor=green,
            citecolor=blue
            ]{hyperref}

\begin{document}

\preprint{AIP/123-QED}

\title[Sample title]{Surface coating with oxide layers to enhance the spin properties of shallow NV centers in diamond}
\affiliation{ 
Key Laboratory of Strongly-Coupled Matter Physics, Chinese Academy of Sciences, Hefei National Laboratory for Physical Science at Microscale, and Department of Physics, University of Science and Technology of China. Hefei, Anhui, 230026, P.R.China.
}%
\author{Wenlong Zhang, Shengran Lin, Jian Zhang, Jiaxin Zhao, Yuanjie Yang, Changfeng Weng, Haolei Zhou, Liren Lou, Wei Zhu}
\author{Guanzhong Wang}
 \email{gzwang@ustc.edu.cn.}

\date{\today}

\begin{abstract}
We present an enhancement of spin properties of the shallow (<5nm) NV centers by using an ALD to deposit titanium oxide layer on the diamond surface. With the oxide layer of an appropriate thickness, increases about 2\ensuremath{\sim}3.5 times of both evolution time ($T_{2}^{*}$) and relaxation time ($T_{1}$) were achieved and the shallow NV center charge states stabilized as well. Moreover, the coherence time ($T_{2}$) kept almost unchanged. This surface coating technique could produce a protective coating layer of controllable thickness without any damages to the solid quantum system surface, making it possible to prolong $T_{2}^{*}$ time and $T_{1}$ time, which might be an approach to the further packaging technique for the applicating solid quantum devices.
\end{abstract}

\maketitle
The nitrogen-vacancy (NV) center in diamond has been considered to be prospective
in both quantum information science\cite{jelezko2006single} and precision measurement
\cite{schirhagl2014nitrogen} owing to its outstanding spin properties 
at room temperature. For example, quantum registers based on the coupling of two 
NV centers have been experimentally demonstrated.\cite{neumann2010quantum,dolde2013room}  
Measuring techniques of electric\cite{dolde2011electric,kim2015decoherence} 
and magnetic fields,\cite{steinert2010high,Grinolds2013Nanoscale,
rosskopf2014investigation,rondin2014magnetometry,jarmola2012temperature} temperature\cite{jarmola2012temperature,neumann2013high,kucsko2013nanometre,wang2015high} 
etc., by using NV centers in diamond have driven to maturity stage. It is of interest
to make the solid quantum system of diamond NV center a device for application. 
However, in experimental quantum information science, one challenge is to insulate the
quantum systems from the external interferences. Up to now,
a number of experimental schemes to relieve the influences from   
diamond surface or outer environment have been presented and tested by many groups.\cite{hauf2011chemical,grotz2012charge,kaviani2014proper,kageura2017effect,loretz2014nanoscale,
Kim2014Effect,Wang2016Coherence,cui2015reduced,favaro2015effect,Zhang2017Depth,staudacher2012enhancing}
One familiar and effective method is to make the diamond surface 
$O$-terminated or $N$-terminated etc., called surface termination.
\cite{hauf2011chemical,grotz2012charge,kaviani2014proper,kageura2017effect} However, 
this surface treatment can provide only an atom-thick layer on the diamond surface 
and the limited layer thickness may not meet the needs of packaging technique. 
Moreover, thermal oxidation\cite{loretz2014nanoscale,Kim2014Effect,Wang2016Coherence} or plasma etching\cite{cui2015reduced,favaro2015effect,Zhang2017Depth} 
used to obtain the satisfactory surface termination damages the diamond surface irreversibly, 
etching away the very shallow (<5nm) NV centers. Hence a method which can produce a protective 
layer of controllable thickness without any surface damages to prevent from 
etching shallow centers away is being sought.

In this letter, we demonstrate an enhancement of the spin properties of shallow 
NV centers by depositing titanium oxide coating layer on the diamond surface without any surface damages. 
We performed atomic layer deposition (ALD) to deposit nanoscale titanium oxide which is nonmagnetic and
non-fluorescent at room temperature as surface coating layer.
Before and after ALD, we studied the variations of shallow center spin 
properties, such as free induced evolution time ($T_{2}^{*}$), 
coherence time ($T_{2}$) and spin-lattice relaxation time ($T_{1}$),
the behaviors being frequently-used in various applications. We also studied 
the relevance between spin properties and layer thickness. 
This surface coating technique may contribute to the prospective packaging 
technique for devices made of diamond NV center systems.

An array comprising 60 nm diameter apertures 
was patterned on a 300-nm-thick polymethyl methacrylate (PMMA) layer deposited 
beforehand on the surface of an electronic grade diamond substrate from $Element$ $Six$.\cite{wang2015high,spinicelli2011engineered}
By ion implantation with the $\textsuperscript{14}N\textsubscript{2}\textsuperscript{+}$
molecule energy of 50 keV and a fluence of 0.65 \texttimes{} 10\textsuperscript{11}
$\textsuperscript{14}N\textsubscript{2}\textsuperscript{+}$ per cm\textsuperscript{2}
through the apertures, the NV center array was created with a probable depth of about 33 nm 
(obtained by SRIM). Afterwards, the implanted diamond substrate was annealed at 1050 \textcelsius{}
in vacuum at 2 \texttimes{} 10 \textsuperscript{-5} Pa for 2 h, 
and then boiled in a 1:1:1 mixture of sulfuric, nitric, and perchloric acids. 
The fluorescence image of a representative area of NV center array after all the processes 
above implemented is shown in Fig.1(a) where the lightspots represent NV centers and 
the fluorescence intensity of a single NV (verified by its $g^{(2)}(0)$ value much 
less than 0.5) is about 30k /s.

\begin{figure}[!h]
\centering
\includegraphics[height=11.8cm,width=8.6cm]{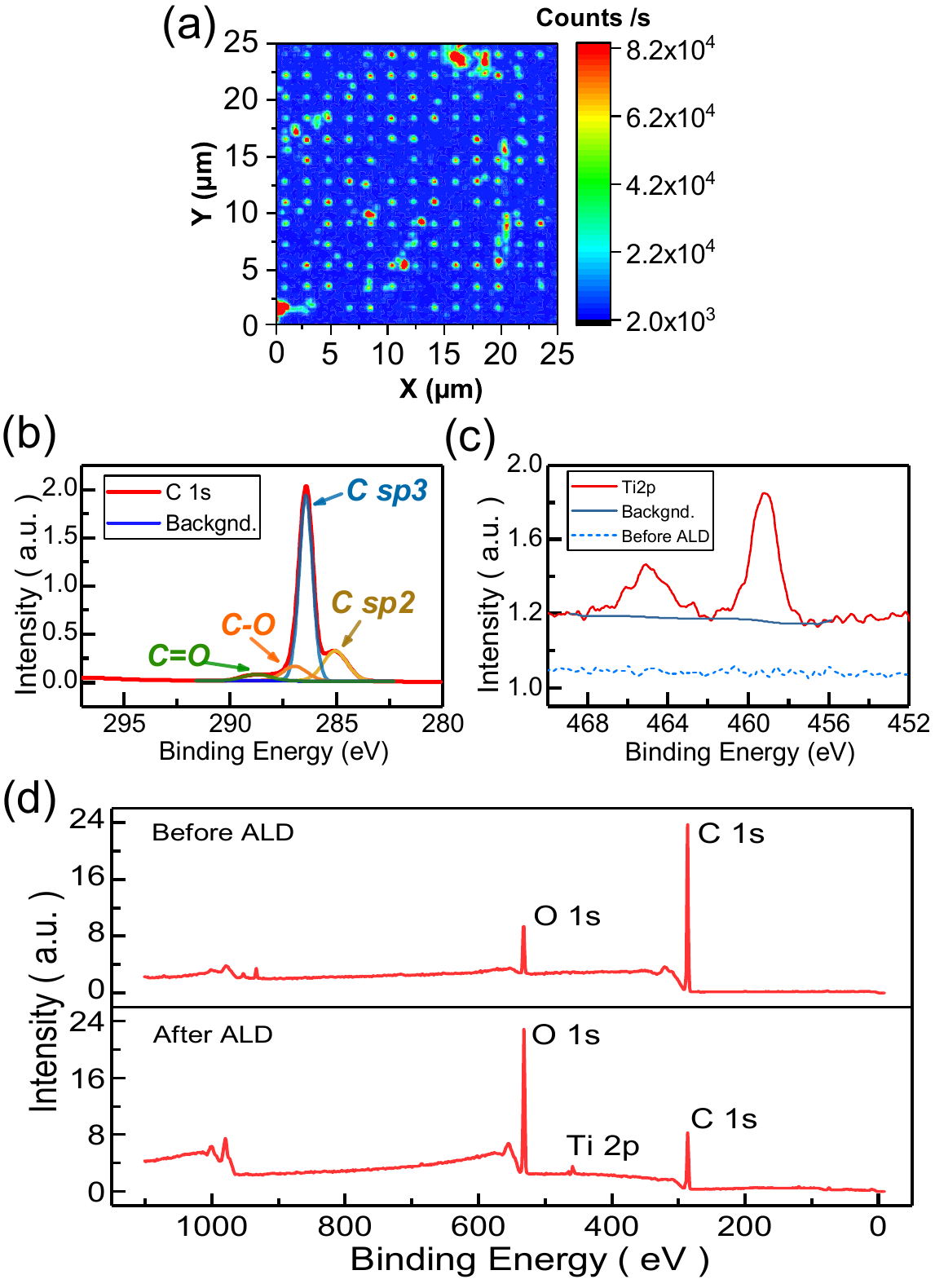}
\caption{\label{fig:epsart} 
(a) The fluorescence image of a representative area of NV center array in the implanted diamond. (b) High-resolution XPS carbon $1s$ spectrum of the diamond sample surface. (c) High-resolution XPS titanium $2p$ spectra of the diamond sample surface. The dotted line shows the result before ALD and the solid line shows the result after ALD. (d) XPS survey scan showing carbon, oxygen and titanium peaks, before ALD [top] and after ALD [bottom], respectively.}
\end{figure}

To obtain shallow centers, we performed plasma etching by 
using a reactive-ion-etch (RIE) reactor (Oxford PlasmaPro NGP80 machine).
The diamond plate was etched in conditions of 200 w ICP power, 30
mTorr chamber pressure, 10 sccm of oxygen, 5 sccm of argon, with an
etching rate of 11.4 \textpm{} 2 nm/min.\cite{Zhang2017Depth} We applied a 150-second plasma
etching on the diamond (corresponding to an etching depth of about
28.5 nm) to make most NVs locate at the depths less than 5 nm from the surface.
Using the shallow centers obtained by plasma etching rather than 
using the ready-made shallow centers generated by low energy implantation 
was because we could contrast the etched shallow center behaviours 
after ALD with the ones of these centers locating deep in the 
diamond before etching to verify the effects of surface coating. 
On the other hand, the first cycle of $H_{2}O$ in the deposition processes 
would make the diamond surface $O$-terminated, so we made the sample 
surface $O$-terminated by oxygen plasma etching before surface coating 
to eliminate the contribution of oxygen termination to the  
spin property variations in contrastive study.  

\begin{figure}
\centering
\includegraphics[height=8.4cm,width=6cm]{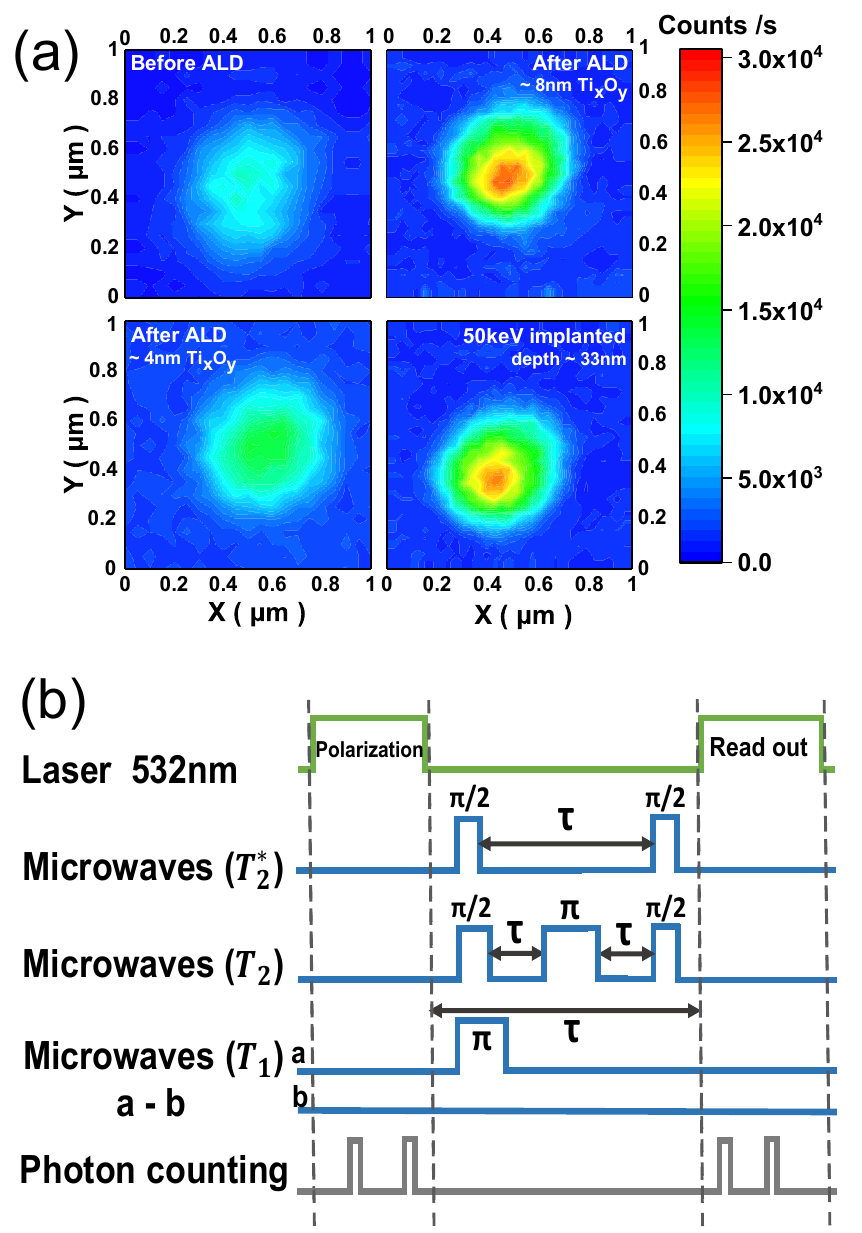}
\caption{\label{fig:epsart} 
(a) the fluorescence images of the same single center (NV-02) after
ion implantation [bottom-right], after plasma etching [top-left], after the first ALD [bottom-left] 
and after the second ALD [top-right], respectively. (b) Microwave pulse sequences. From the
top down: laser control, $T_{2}^{*}$ measurement sequence, $T_{2}$ measurement sequence, 
$T_{1}$ measurement sequence and photon counting as the background
reference.}
\end{figure}

The diamond sample was then put into the ALD chamber (Picosun, Sunale R-200 Advanced) 
to deposit titanium oxide layer ($Ti_{x}O_{y}$ layer, abbreviated as TOL hereafter). The first atomic 
deposition cycle of $H_{2}O$ cycle would make the hydroxyl ions absorb on the sample surface. 
Then the titanium ions from the second deposition $Ti$ cycle would replace the hydrogen 
ions of hydroxy. We repeated the above two cycles sequentially to 
deposit the oxide layer on the diamond surface (An additional $H_{2}O$ 
cycle was always applied in the end of all ALD processes.). The deposition rate of TOL 
was about 1 nm for every nine-ALD cycle at the temperature of 120 \textcelsius.
Before and after ALD, the sample surface was characterized with 
X-ray photoelectron spectroscopy (XPS) to verify if the
TOL had been deposited on the surface.

\begin{figure*}
\centering
\includegraphics[height=9.6cm,width=17.5cm]{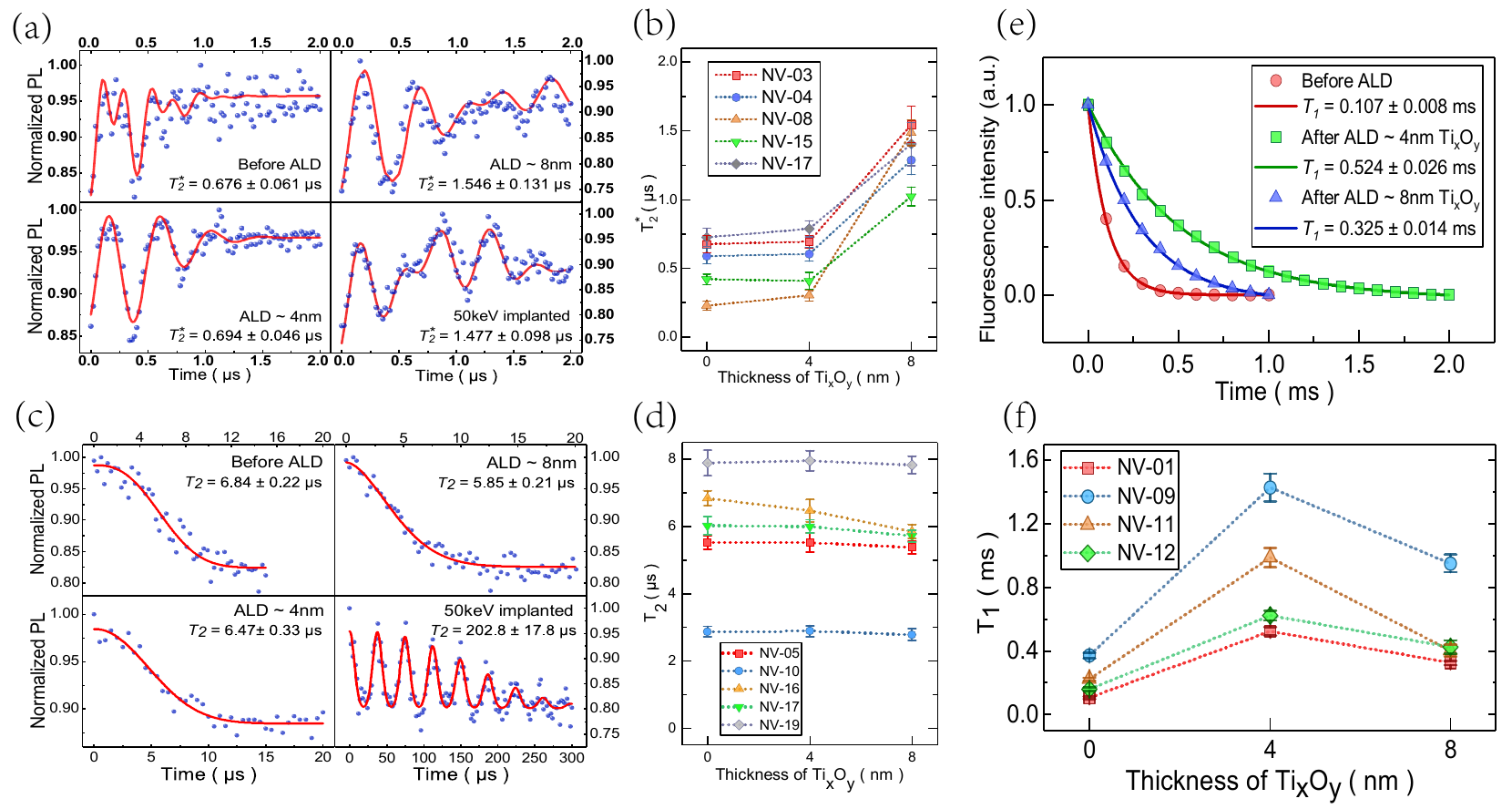}
\caption{\label{fig.epsart}
$T_{2}^{*}$, $T_{2}$ and $T_{1}$ measurements of the labeled shallow single NV centers before and after ALD. 
(a) The Ramsey measurements of NV-03 before ALD (top-left), after depositing
4 nm TOL (bottom-left), after depositing 8 nm TOL (top-right),
and initially at the depth about 33 nm (bottom-right), 
respectively. (b) The Ramsey measurements of five typical centers before ALD, 
after depositing 4 nm TOL and after depositing 8 nm TOL. (c) The spin echo measurements 
of NV-16 before ALD (top-left), after depositing 4 nm TOL (bottom-left), after 
depositing 8 nm TOL (top-right), and initially at the depth about 33 nm (bottom-right), 
respectively. (d) The spin encho measurements of five typical single centers 
before ALD, after depositing 4 nm TOL and after depositing 8 nm TOL. (e) $T_{1}$ measurements 
of NV-01 before ALD, after depositing 4 nm TOL and after depositing 8 nm TOL. (f) $T_{1}$
measurements of four typical single NV centers before ALD,
after depositing 4 nm TOL and after depositing 8 nm TOL.}
\end{figure*}

Figure 1(b) exhibits the high-resolution XPS carbon $1s$ spectrum 
of the sample surface after ALD, showing a dominant diamond $sp^{3}$ peak at 286.5 eV, 
accompanied with three side peaks, from left to right, assigned 
to carbon doubly and singly-bonded to oxygen, and $sp^{2}$ carbon. 
Comparing Fig.1(b) with the XPS result of carbon $1s$ spectrum of the same sample  
before ALD (not shown in this letter), the two spectra were almost identical.
Figure 1(c) presents the high-resolution XPS titanium $2p$ spectra. 
The dotted line shows the result before ALD from which no titanium related peaks 
are obtained, while the solid line represents the result after ALD from which 
distinct peaks belonging to $Ti$ $2p$ spectrum were found around 459\ eV. 
Another proof to certify the success of depositing oxide layer on the diamond 
surface was the XPS survey scan exhibited in Fig.1(d), showing the relative intensity  
of oxygen and carbon peaks. The intensity of oxygen peak raised remarkably relative to 
the intensity of carbon peak, revealing that after ALD, the main component of characterized 
sample surface changed from carbon to oxygen, i.e., the oxide layer had been indeed 
deposited on the surface. Moreover, the appearance of $Ti$ peaks in 
Fig.1(d, bottom) demonstrated that the protective layer was made up of titanium oxide.

The diamond sample was placed in the custom-built confocal microscope
system with the applied magnetic field (B = 55 \textpm{} 5 G) paralleling
to the detected single center axes. The ALD was performed
twice in succession to study the influence of TOL thickness on spin properties. 
Twenty shallow single NV centers (labeled NV-01\textasciitilde{}20) were 
measured before and after each ALD. Figure 2(a) demonstrates
the fluorescence images of the same single center (NV-02)
after each step. This representative center initially sat at the depth
of about 33 nm with its photoluminescence (PL) intensity of about
28k /s [Fig.2(a) bottom-right]. After proper etching treatment NV-02 
became close to the surface with its depth less than
5 nm and its PL intensity decreased to less
than 10k /s [Fig.2(a) top-left] owing to the unstable 
charge state of shallow center.\cite{schirhagl2014nitrogen,bradac2010observation,Newell2016Surface} 
When a 4-nm-thick TOL was deposited on the diamond surface, the PL intensity of NV-02 
increased to about 15k /s [Fig.2(a) bottom-left], indicating that its charge state became
stable. Moreover, after another 4-nm-thick TOL being deposited (8 nm in total), 
the NV-02's PL intensity increased to about 28k /s [Fig.2(a) top-right],  
similar to its initial PL intensity. For all the measured single centers, 
their PL intensities changed in the same way. As stated above, the TOL in our 
experiment was nanoscale so that the optical collection efficiency of 
objective was unchanged. Therefore, the increase of NV center's PL intensity 
demonstrated that the charge states of shallow NVs could 
stabilize by surface coating titanium oxide protective layer.

The variations of free induced evolution ($T_{2}^{*}$), coherence ($T_{2}$)
and spin-lattice relaxation ($T_{1}$) arising from the deposited TOL were studied by 
applying basic microwave (MW) pulse sequences [Fig.2(b)]. 
The oscillating curves of evolution times were fitted with the function
\[
I(t)=I_{0}+A_{0}\exp[-(t/T_{2}^{*})^{n}]\sum_{i=1}^{3}\cos(2\pi\delta_{i}t+\phi_{i})
\]
to deduce the spin dephasing time $T_{2}^{*}$, where $\delta_{i}$ is the
detuning frequency and the exponential index n is set to be 2 corresponding
to the Gaussian dephasing case. To acquire the values of $T_{2}$ times, we fitted the data 
by a stretched exponential envelope multiplied by a periodic Gaussian function 
\[
C(t)=C_{0}+A_{0}\exp\Biggl[-\biggl(\frac{t}{T_{2}}\biggr)^{n}\Biggr]\times\underset{i}{\sum}\exp\Biggl[-\biggl(\frac{t-i\times T_{R}}{T_{C}}\biggr)^{2}\Biggr]
\]
where $T\textsubscript{2}$ is the decay time of the envelope of revival peaks, $T\textsubscript{C}$
is the decay time of initial collapse, and $T\textsubscript{R}$
is the revival period time of revival peaks. The $T_{1}$ times were obtained
by fitting data with the single-exponential function $\triangle_{Fl}(t)=A\exp(-t/T_{1})$.
 
The $T_{2}^{*}$ times of a typical center NV-03 measured after different surface 
treatment steps were displayed in Fig.3(a). The different waveforms of Ramsey
fringes are caused by various MW frequence detunings, which does not influence 
the fitting values of $T_{2}^{*}$.\cite{hu2012influence} 
There was a pimping increase of $T_{2,4nm}^{*}$
for NV-03 with a 4-nm-thick TOL on the surface, compared
with its $T_{2,0nm}^{*}$ before ALD. Another deposition
was applied to make the total TOL thickness to be 8 nm and the $T_{2,8nm}^{*}$ 
of NV-03 increased to 1.546 \textpm{} 0.131 $\mu$s,
about 2.5 times of $T_{2,0nm}^{*}$ (0.676 \textpm{} 0.061 $\mu$s).
The value of $T_{2,8nm}^{*}$ was found even larger than
that of $T_{2,50keV}^{*}$ for NV-03 just after being implanted. 
For all measured NVs, the variations of their $T_{2}^{*}$ times with the TOL thickness 
were analogous. Five typical centers' $T_{2}^{*}$ data are exhibited in Fig.3(b), showing
prominent increases in $T_{2}^{*}$ times after the second deposition.

As presented in Fig.3(c), the coherence time $T_{2,4nm}$ of NV-16 was almost unchanged 
after the first ALD and then reduced to, after the second ALD, about eighty 
percents of $T_{2,0nm}$ before ALD. Evidently, $T_{2,8nm}$ (5.85 \textpm{} 0.21 $\mu$s) after surface coating could not turn back to the magnitude of $T_{2,50keV}$ (202.8 \textpm{} 17.8 $\mu$s) for the center just after being implanted. Among all the 20 NVs, the change of $T_{2}$ time of NV-16 
was the most obvious one, exhibited together with other four typical 
centers in Fig.3(d), from which we could find that $T_{2}$ times of other NVs remained
almost the same values before and after each ALD. In other words, $T_{2}$ time could 
keep almost unchanged for most shallow NVs after surface coating.

\begin{figure}
\includegraphics[height=7.5cm,width=6.2cm]{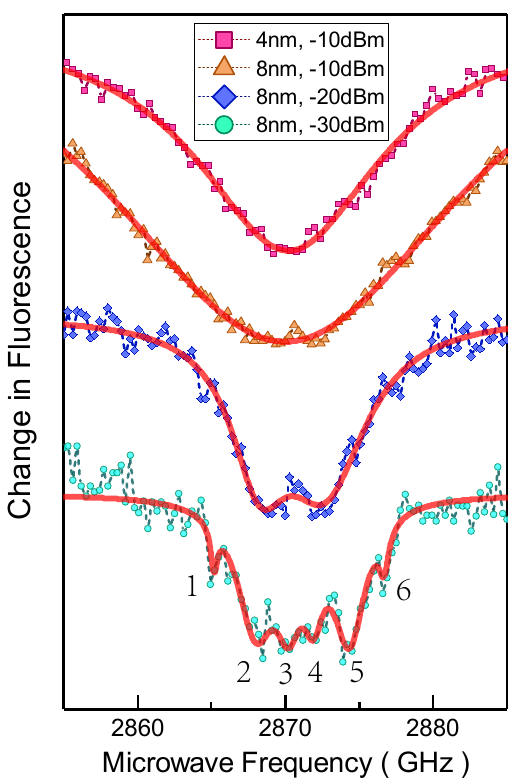}
\caption{\label{fig.epsart}
Zero field CW-ODMR spectra of the same single NV center
with microwave power -10 dBm after depositing 4 nm TOL and -10 dBm,
-20 dBm, -30 dBm after depositing 8 nm TOL, from top down in order
respectively , the solid curves is the Lorentzian fitting to the experimental spectra.
}
\end{figure}

For the longitudinal spin relaxation time $T_{1}$, a representative result of 
the center NV-01 is demonstrated in Fig.3(e). After etching the sample surface but with no
TOL on it, this shallow center's $T_{1,0nm}$ time was deduced to
be about 0.107 \textpm{} 0.008 ms. After the 4-nm-thick TOL was deposited, 
the value of $T_{1,4nm}$ increased to 0.524 \textpm{} 0.026 ms by a factor of about 5. However, when
the TOL was thickened to 8 nm, the $T_{1,8nm}$ of NV-01 dropped to 0.325
\textpm{} 0.014 ms. The variations of all 20 measured
centers' $T_{1}$ times conformed to the rule: $T_{1,4nm}\thickapprox4\thicksim5T_{1,0nm}$, $T_{1,8nm}\thickapprox2\thicksim3.5T_{1,0nm}$, that is $T_{1,0nm}<T_{1,8nm}<T_{1,4nm}$  
as demonstrated in Fig.4(f) for four typical centers.

A number of prior works have investigated the various factors affecting the spin properties. 
It is acknowledged that $T_{2}$ measured by spin echo 
is governed by low-frequency (\ensuremath{\sim}kHz) noises and $T_{2}^{*}$ can be 
influenced by the nosies of even lower frequency ($<$kHz), while $T_{1}$ is 
sensitive to high-frequncy noises.\cite{Myers2014Probing,romach2015spectroscopy,rosskopf2014investigation} 
For the first ALD, that the TOL could be deposited on the diamond
surface indicated that the $O$-terminated surface obtained by
oxygen plasma etching was incomplete, leaving a few unpaired electrons (dangling bonds), 
hydrogen and magnetic ions on the surface. These electrons and ions were connected to or
displaced by titanium or oxygen atoms during ALD. The elimination of surface
spin baths led to the reduction of very low frequency ($<$kHz) nosies, so $T_{2}^{*}$ times 
increased slightly. However, the low-frequency (\ensuremath{\sim}kHz) noises were not related 
much to the surface,\cite{romach2015spectroscopy,rosskopf2014investigation} 
so the first ALD did not influence $T_{2}$ time. The depth dependent 
decoherence was considered to be connected with the magnetic and electric fields 
produced by surface spins previously,\cite{Myers2014Probing,Wang2016Coherence,Zhang2017Depth,kim2015decoherence} 
nevertheless, the main source of external fields 
should be more, in our assumption, attributed to the polar bonds like C-O bonds on the 
diamond surface since the interfacial fields were nearly
unchanged before and after the first ALD as indicated by the almost unchanged $T_{2}$ times.
As for the increase in $T_{1}$ time, the decrease of magnetic ions absorbing on the diamond 
surface and the reduction of surface effects would be responsible.\cite{staudacher2012enhancing,romach2015spectroscopy,rosskopf2014investigation}

As the TOL was thickened from 4 to 8 nm, $T_{1}$ reduced while $T_{2}^{*}$ 
sharply increased. To understand this, we compared the two ODMR spectra of the same center
with 4 and 8 nm TOL on the surface without any applied magnetic fields. Although both ODMR
spectra were obtained in the same measurement condition
of -10 dBm MW power, an inappreciable dip around 2.87
GHz was found in the spectrum in the case of 8-nm-thick TOL 
[triangle symbols] shown in Fig.(4). To eliminate the line broadening
existent under strong applied fields, we lowered the MW power to -20
dBm and the dip [rhombus symbols] became distinct. Lowering again the MW power to -30 dBm, the
ODMR spectrum [circular symbols] showed a split around 2.87
GHz. The curves were obtained by using Lorentz fit and six peaks could be acquired. 
Except for the interval between peak-3 and peak-4, all the other intervals
between two adjacent peaks were about 2.2 MHz, coinciding with the
hyperfine coupling constant of $^{14}N$.\cite{rabeau2006implantation} 
It was well-known that the hyperfine splitting
of $^{14}N$ would produce three peaks and that here
it ulteriorly divided into two groups of six peaks was owing
to the existense of additional electric or strain field. This additional field 
was introduced from the external stress generated by lattice mismatch of 
the unbroken expanse of TOL with the intensity of about 6 MHz as derived 
from the fitting curve, leading to the reduction of both $T_{1}$ and $T_{2}$ times.
Although $T_{1,8nm}$ decreased after the second ALD, it was still longer 
than $T_{1,0nm}$ of the very shallow NVs without surface coating, i.e., 
the spin property was generally enhanced. Moreover, it has been reported that $T_{2}^{*}$ is enormously
influenced by inhomogeneous magnetic felds while the external stress
can suppress this influence to prolong the $T_{2}^{*}$.\cite{zhou2017self} 
Therefore, the different behaviors of spin properties were due to the 
appearance of external stress. In addition, we found a few centers' 
$T_{2,8nm}^{*}$ were even larger than their $T_{2,50keV}^{*}$ like NV-03 shown in Fig.3(a), suggesting
that a part of the inhomogeneous magnetic felds were attributed
to the applied magnetic field by using alternating electromagnet.

Consequently, using ALD technique to deposit titanium oxide layer can enhance 
the spin properties of shallow NV centers in diamond with an increase about 
2\textasciitilde{}3.5 times for both relaxation time ($T_{1}$) and evolution time ($T_{2}^{*}$) 
with almost unchanged coherence time ($T_{2}$). This surface coating produces 
a protective layer of controllable thickness on the surface of solid quantum system 
like diamond NV center without any surface damages. Furthermore, 
this technique may also provide a possible approach for packaging technique for
diamond NV center devices used in quantum information science 
and precision measurement.\\

This work was supported by the National Natural Science Foundation of 
China (Grant No.11374280 and No.50772110). The authors wish to thank G.P.Guo, J.You and Y.Li
from the Key lab of Quantum Information for the support of electron
beam lithography. We also thank M.L.Li from the Department of Physics, University of Science and Technology of China, for the support of ALD technique. We further thank X.X.Wang, J.L.Peng, D.F.Zhou and W.Liu from the USTC Center for Micro- and Nanoscale Research and Fabrication for the technical
support of Plasma etching. \\

\bibliographystyle{rsc}

\begin{mcitethebibliography}{39}
\providecommand*{\natexlab}[1]{#1}
\providecommand*{\mciteSetBstSublistMode}[1]{}
\providecommand*{\mciteSetBstMaxWidthForm}[2]{}
\providecommand*{\mciteBstWouldAddEndPuncttrue}
  {\def\EndOfBibitem{\unskip.}}
\providecommand*{\mciteBstWouldAddEndPunctfalse}
  {\let\EndOfBibitem\relax}
\providecommand*{\mciteSetBstMidEndSepPunct}[3]{}
\providecommand*{\mciteSetBstSublistLabelBeginEnd}[3]{}
\providecommand*{\EndOfBibitem}{}
\mciteSetBstSublistMode{f}
\mciteSetBstMaxWidthForm{subitem}
{(\emph{\alph{mcitesubitemcount}})}
\mciteSetBstSublistLabelBeginEnd{\mcitemaxwidthsubitemform\space}
{\relax}{\relax}

\bibitem[Jelezko and Wrachtrup(2006)]{jelezko2006single}
F.~Jelezko and J.~Wrachtrup, \emph{physica status solidi (a)}, 2006,
  \textbf{203}, 3207--3225\relax
\mciteBstWouldAddEndPuncttrue
\mciteSetBstMidEndSepPunct{\mcitedefaultmidpunct}
{\mcitedefaultendpunct}{\mcitedefaultseppunct}\relax
\EndOfBibitem
\bibitem[Schirhagl \emph{et~al.}(2014)Schirhagl, Chang, Loretz, and
  Degen]{schirhagl2014nitrogen}
R.~Schirhagl, K.~Chang, M.~Loretz and C.~L. Degen, \emph{Annual review of
  physical chemistry}, 2014, \textbf{65}, 83--105\relax
\mciteBstWouldAddEndPuncttrue
\mciteSetBstMidEndSepPunct{\mcitedefaultmidpunct}
{\mcitedefaultendpunct}{\mcitedefaultseppunct}\relax
\EndOfBibitem
\bibitem[Neumann \emph{et~al.}(2010)Neumann, Kolesov, Naydenov, Beck, Rempp,
  Steiner, Jacques, Balasubramanian, Markham,
  Twitchen,\emph{et~al.}]{neumann2010quantum}
P.~Neumann, R.~Kolesov, B.~Naydenov, J.~Beck, F.~Rempp, M.~Steiner, V.~Jacques,
  G.~Balasubramanian, M.~Markham, D.~Twitchen \emph{et~al.}, \emph{Nature
  Physics}, 2010, \textbf{6}, 249\relax
\mciteBstWouldAddEndPuncttrue
\mciteSetBstMidEndSepPunct{\mcitedefaultmidpunct}
{\mcitedefaultendpunct}{\mcitedefaultseppunct}\relax
\EndOfBibitem
\bibitem[Dolde \emph{et~al.}(2013)Dolde, Jakobi, Naydenov, Zhao, Pezzagna,
  Trautmann, Meijer, Neumann, Jelezko, and Wrachtrup]{dolde2013room}
F.~Dolde, I.~Jakobi, B.~Naydenov, N.~Zhao, S.~Pezzagna, C.~Trautmann,
  J.~Meijer, P.~Neumann, F.~Jelezko and J.~Wrachtrup, \emph{Nature Physics},
  2013, \textbf{9}, 139\relax
\mciteBstWouldAddEndPuncttrue
\mciteSetBstMidEndSepPunct{\mcitedefaultmidpunct}
{\mcitedefaultendpunct}{\mcitedefaultseppunct}\relax
\EndOfBibitem
\bibitem[Dolde \emph{et~al.}(2011)Dolde, Fedder, Doherty, N{\"o}bauer, Rempp,
  Balasubramanian, Wolf, Reinhard, Hollenberg,
  Jelezko,\emph{et~al.}]{dolde2011electric}
F.~Dolde, H.~Fedder, M.~W. Doherty, T.~N{\"o}bauer, F.~Rempp,
  G.~Balasubramanian, T.~Wolf, F.~Reinhard, L.~C. Hollenberg, F.~Jelezko
  \emph{et~al.}, \emph{Nature Physics}, 2011, \textbf{7}, 459\relax
\mciteBstWouldAddEndPuncttrue
\mciteSetBstMidEndSepPunct{\mcitedefaultmidpunct}
{\mcitedefaultendpunct}{\mcitedefaultseppunct}\relax
\EndOfBibitem
\bibitem[Kim \emph{et~al.}(2015)Kim, Mamin, Sherwood, Ohno, Awschalom, and
  Rugar]{kim2015decoherence}
M.~Kim, H.~Mamin, M.~Sherwood, K.~Ohno, D.~Awschalom and D.~Rugar,
  \emph{Physical review letters}, 2015, \textbf{115}, 087602\relax
\mciteBstWouldAddEndPuncttrue
\mciteSetBstMidEndSepPunct{\mcitedefaultmidpunct}
{\mcitedefaultendpunct}{\mcitedefaultseppunct}\relax
\EndOfBibitem
\bibitem[Steinert \emph{et~al.}(2010)Steinert, Dolde, Neumann, Aird, Naydenov,
  Balasubramanian, Jelezko, and Wrachtrup]{steinert2010high}
S.~Steinert, F.~Dolde, P.~Neumann, A.~Aird, B.~Naydenov, G.~Balasubramanian,
  F.~Jelezko and J.~Wrachtrup, \emph{Review of scientific instruments}, 2010,
  \textbf{81}, 043705\relax
\mciteBstWouldAddEndPuncttrue
\mciteSetBstMidEndSepPunct{\mcitedefaultmidpunct}
{\mcitedefaultendpunct}{\mcitedefaultseppunct}\relax
\EndOfBibitem
\bibitem[Grinolds \emph{et~al.}(2013)Grinolds, Hong, Maletinsky, Luan, Lukin,
  Walsworth, and Yacoby]{Grinolds2013Nanoscale}
M.~S. Grinolds, S.~Hong, P.~Maletinsky, L.~Luan, M.~D. Lukin, R.~L. Walsworth
  and A.~Yacoby, \emph{Nature Physics}, 2013, \textbf{9}, 215--219\relax
\mciteBstWouldAddEndPuncttrue
\mciteSetBstMidEndSepPunct{\mcitedefaultmidpunct}
{\mcitedefaultendpunct}{\mcitedefaultseppunct}\relax
\EndOfBibitem
\bibitem[Rosskopf \emph{et~al.}(2014)Rosskopf, Dussaux, Ohashi, Loretz,
  Schirhagl, Watanabe, Shikata, Itoh, and Degen]{rosskopf2014investigation}
T.~Rosskopf, A.~Dussaux, K.~Ohashi, M.~Loretz, R.~Schirhagl, H.~Watanabe,
  S.~Shikata, K.~Itoh and C.~Degen, \emph{Physical review letters}, 2014,
  \textbf{112}, 147602\relax
\mciteBstWouldAddEndPuncttrue
\mciteSetBstMidEndSepPunct{\mcitedefaultmidpunct}
{\mcitedefaultendpunct}{\mcitedefaultseppunct}\relax
\EndOfBibitem
\bibitem[Rondin \emph{et~al.}(2014)Rondin, Tetienne, Hingant, Roch, Maletinsky,
  and Jacques]{rondin2014magnetometry}
L.~Rondin, J.~Tetienne, T.~Hingant, J.~Roch, P.~Maletinsky and V.~Jacques,
  \emph{Reports on progress in physics}, 2014, \textbf{77}, 056503\relax
\mciteBstWouldAddEndPuncttrue
\mciteSetBstMidEndSepPunct{\mcitedefaultmidpunct}
{\mcitedefaultendpunct}{\mcitedefaultseppunct}\relax
\EndOfBibitem
\bibitem[Jarmola \emph{et~al.}(2012)Jarmola, Acosta, Jensen, Chemerisov, and
  Budker]{jarmola2012temperature}
A.~Jarmola, V.~Acosta, K.~Jensen, S.~Chemerisov and D.~Budker, \emph{Physical
  review letters}, 2012, \textbf{108}, 197601\relax
\mciteBstWouldAddEndPuncttrue
\mciteSetBstMidEndSepPunct{\mcitedefaultmidpunct}
{\mcitedefaultendpunct}{\mcitedefaultseppunct}\relax
\EndOfBibitem
\bibitem[Neumann \emph{et~al.}(2013)Neumann, Jakobi, Dolde, Burk, Reuter,
  Waldherr, Honert, Wolf, Brunner, Shim,\emph{et~al.}]{neumann2013high}
P.~Neumann, I.~Jakobi, F.~Dolde, C.~Burk, R.~Reuter, G.~Waldherr, J.~Honert,
  T.~Wolf, A.~Brunner, J.~H. Shim \emph{et~al.}, \emph{Nano letters}, 2013,
  \textbf{13}, 2738--2742\relax
\mciteBstWouldAddEndPuncttrue
\mciteSetBstMidEndSepPunct{\mcitedefaultmidpunct}
{\mcitedefaultendpunct}{\mcitedefaultseppunct}\relax
\EndOfBibitem
\bibitem[Kucsko \emph{et~al.}(2013)Kucsko, Maurer, Yao, Kubo, Noh, Lo, Park,
  and Lukin]{kucsko2013nanometre}
G.~Kucsko, P.~Maurer, N.~Y. Yao, M.~Kubo, H.~Noh, P.~Lo, H.~Park and M.~D.
  Lukin, \emph{Nature}, 2013, \textbf{500}, 54\relax
\mciteBstWouldAddEndPuncttrue
\mciteSetBstMidEndSepPunct{\mcitedefaultmidpunct}
{\mcitedefaultendpunct}{\mcitedefaultseppunct}\relax
\EndOfBibitem
\bibitem[Wang \emph{et~al.}(2015)Wang, Feng, Zhang, Chen, Zheng, Guo, Zhang,
  Song, Guo, Fan,\emph{et~al.}]{wang2015high}
J.~Wang, F.~Feng, J.~Zhang, J.~Chen, Z.~Zheng, L.~Guo, W.~Zhang, X.~Song,
  G.~Guo, L.~Fan \emph{et~al.}, \emph{Physical Review B}, 2015, \textbf{91},
  155404\relax
\mciteBstWouldAddEndPuncttrue
\mciteSetBstMidEndSepPunct{\mcitedefaultmidpunct}
{\mcitedefaultendpunct}{\mcitedefaultseppunct}\relax
\EndOfBibitem
\bibitem[Hauf \emph{et~al.}(2011)Hauf, Grotz, Naydenov, Dankerl, Pezzagna,
  Meijer, Jelezko, Wrachtrup, Stutzmann,
  Reinhard,\emph{et~al.}]{hauf2011chemical}
M.~Hauf, B.~Grotz, B.~Naydenov, M.~Dankerl, S.~Pezzagna, J.~Meijer, F.~Jelezko,
  J.~Wrachtrup, M.~Stutzmann, F.~Reinhard \emph{et~al.}, \emph{Physical Review
  B}, 2011, \textbf{83}, 081304\relax
\mciteBstWouldAddEndPuncttrue
\mciteSetBstMidEndSepPunct{\mcitedefaultmidpunct}
{\mcitedefaultendpunct}{\mcitedefaultseppunct}\relax
\EndOfBibitem
\bibitem[Grotz \emph{et~al.}(2012)Grotz, Hauf, Dankerl, Naydenov, Pezzagna,
  Meijer, Jelezko, Wrachtrup, Stutzmann,
  Reinhard,\emph{et~al.}]{grotz2012charge}
B.~Grotz, M.~V. Hauf, M.~Dankerl, B.~Naydenov, S.~Pezzagna, J.~Meijer,
  F.~Jelezko, J.~Wrachtrup, M.~Stutzmann, F.~Reinhard \emph{et~al.},
  \emph{Nature communications}, 2012, \textbf{3}, 729\relax
\mciteBstWouldAddEndPuncttrue
\mciteSetBstMidEndSepPunct{\mcitedefaultmidpunct}
{\mcitedefaultendpunct}{\mcitedefaultseppunct}\relax
\EndOfBibitem
\bibitem[Kaviani \emph{et~al.}(2014)Kaviani, Deák, Aradi, Frauenheim, Chou,
  and Gali]{kaviani2014proper}
M.~Kaviani, P.~Deák, B.~Aradi, T.~Frauenheim, J.-P. Chou and A.~Gali,
  \emph{Nano letters}, 2014, \textbf{14}, 4772--4777\relax
\mciteBstWouldAddEndPuncttrue
\mciteSetBstMidEndSepPunct{\mcitedefaultmidpunct}
{\mcitedefaultendpunct}{\mcitedefaultseppunct}\relax
\EndOfBibitem
\bibitem[Kageura \emph{et~al.}(2017)Kageura, Kato, Yamano, Suaebah, Kajiya,
  Kawai, Inaba, Tanii, Haruyama, Yamada,\emph{et~al.}]{kageura2017effect}
T.~Kageura, K.~Kato, H.~Yamano, E.~Suaebah, M.~Kajiya, S.~Kawai, M.~Inaba,
  T.~Tanii, M.~Haruyama, K.~Yamada \emph{et~al.}, \emph{Applied Physics
  Express}, 2017, \textbf{10}, 055503\relax
\mciteBstWouldAddEndPuncttrue
\mciteSetBstMidEndSepPunct{\mcitedefaultmidpunct}
{\mcitedefaultendpunct}{\mcitedefaultseppunct}\relax
\EndOfBibitem
\bibitem[Loretz \emph{et~al.}(2014)Loretz, Pezzagna, Meijer, and
  Degen]{loretz2014nanoscale}
M.~Loretz, S.~Pezzagna, J.~Meijer and C.~Degen, \emph{Applied Physics Letters},
  2014, \textbf{104}, 033102\relax
\mciteBstWouldAddEndPuncttrue
\mciteSetBstMidEndSepPunct{\mcitedefaultmidpunct}
{\mcitedefaultendpunct}{\mcitedefaultseppunct}\relax
\EndOfBibitem
\bibitem[Kim \emph{et~al.}(2014)Kim, Mamin, Sherwood, and
  Rettner]{Kim2014Effect}
M.~Kim, H.~J. Mamin, M.~H. Sherwood and C.~T. Rettner, \emph{Applied Physics
  Letters}, 2014, \textbf{105}, 042406--042406--4\relax
\mciteBstWouldAddEndPuncttrue
\mciteSetBstMidEndSepPunct{\mcitedefaultmidpunct}
{\mcitedefaultendpunct}{\mcitedefaultseppunct}\relax
\EndOfBibitem
\bibitem[Wang \emph{et~al.}(2016)Wang, Zhang, Zhang, You, Li, Guo, Feng, Song,
  Lou, and Zhu]{Wang2016Coherence}
J.~Wang, W.~Zhang, J.~Zhang, J.~You, Y.~Li, G.~Guo, F.~Feng, X.~Song, L.~Lou
  and W.~Zhu, \emph{Nanoscale}, 2016, \textbf{8}, 5780--5785\relax
\mciteBstWouldAddEndPuncttrue
\mciteSetBstMidEndSepPunct{\mcitedefaultmidpunct}
{\mcitedefaultendpunct}{\mcitedefaultseppunct}\relax
\EndOfBibitem
\bibitem[Cui \emph{et~al.}(2015)Cui, Greenspon, Ohno, Myers, Jayich, Awschalom,
  and Hu]{cui2015reduced}
S.~Cui, A.~S. Greenspon, K.~Ohno, B.~A. Myers, A.~C.~B. Jayich, D.~D. Awschalom
  and E.~L. Hu, \emph{Nano letters}, 2015, \textbf{15}, 2887--2891\relax
\mciteBstWouldAddEndPuncttrue
\mciteSetBstMidEndSepPunct{\mcitedefaultmidpunct}
{\mcitedefaultendpunct}{\mcitedefaultseppunct}\relax
\EndOfBibitem
\bibitem[F{\'a}varo~de Oliveira \emph{et~al.}(2015)F{\'a}varo~de Oliveira,
  Momenzadeh, Wang, Konuma, Markham, Edmonds, Denisenko, and
  Wrachtrup]{favaro2015effect}
F.~F{\'a}varo~de Oliveira, S.~A. Momenzadeh, Y.~Wang, M.~Konuma, M.~Markham,
  A.~M. Edmonds, A.~Denisenko and J.~Wrachtrup, \emph{Applied Physics Letters},
  2015, \textbf{107}, 073107\relax
\mciteBstWouldAddEndPuncttrue
\mciteSetBstMidEndSepPunct{\mcitedefaultmidpunct}
{\mcitedefaultendpunct}{\mcitedefaultseppunct}\relax
\EndOfBibitem
\bibitem[Zhang \emph{et~al.}(2017)Zhang, Zhang, Wang, Feng, Lin, Lou, Zhu, and
  Wang]{Zhang2017Depth}
W.~Zhang, J.~Zhang, J.~Wang, F.~Feng, S.~Lin, L.~Lou, W.~Zhu and G.~Wang,
  \emph{Physical Review B}, 2017, \textbf{96}, 235443\relax
\mciteBstWouldAddEndPuncttrue
\mciteSetBstMidEndSepPunct{\mcitedefaultmidpunct}
{\mcitedefaultendpunct}{\mcitedefaultseppunct}\relax
\EndOfBibitem
\bibitem[Staudacher \emph{et~al.}(2012)Staudacher, Ziem, H{\"a}ussler,
  St{\"o}hr, Steinert, Reinhard, Scharpf, Denisenko, and
  Wrachtrup]{staudacher2012enhancing}
T.~Staudacher, F.~Ziem, L.~H{\"a}ussler, R.~St{\"o}hr, S.~Steinert,
  F.~Reinhard, J.~Scharpf, A.~Denisenko and J.~Wrachtrup, \emph{Applied Physics
  Letters}, 2012, \textbf{101}, 212401\relax
\mciteBstWouldAddEndPuncttrue
\mciteSetBstMidEndSepPunct{\mcitedefaultmidpunct}
{\mcitedefaultendpunct}{\mcitedefaultseppunct}\relax
\EndOfBibitem
\bibitem[Spinicelli \emph{et~al.}(2011)Spinicelli, Dr{\'e}au, Rondin, Silva,
  Achard, Xavier, Bansropun, Debuisschert, Pezzagna,
  Meijer,\emph{et~al.}]{spinicelli2011engineered}
P.~Spinicelli, A.~Dr{\'e}au, L.~Rondin, F.~Silva, J.~Achard, S.~Xavier,
  S.~Bansropun, T.~Debuisschert, S.~Pezzagna, J.~Meijer \emph{et~al.},
  \emph{New Journal of Physics}, 2011, \textbf{13}, 025014\relax
\mciteBstWouldAddEndPuncttrue
\mciteSetBstMidEndSepPunct{\mcitedefaultmidpunct}
{\mcitedefaultendpunct}{\mcitedefaultseppunct}\relax
\EndOfBibitem
\bibitem[Bradac \emph{et~al.}(2010)Bradac, Gaebel, Naidoo, Sellars, Twamley,
  Brown, Barnard, Plakhotnik, Zvyagin, and Rabeau]{bradac2010observation}
C.~Bradac, T.~Gaebel, N.~Naidoo, M.~Sellars, J.~Twamley, L.~Brown, A.~Barnard,
  T.~Plakhotnik, A.~Zvyagin and J.~Rabeau, \emph{Nature nanotechnology}, 2010,
  \textbf{5}, 345\relax
\mciteBstWouldAddEndPuncttrue
\mciteSetBstMidEndSepPunct{\mcitedefaultmidpunct}
{\mcitedefaultendpunct}{\mcitedefaultseppunct}\relax
\EndOfBibitem
\bibitem[Newell \emph{et~al.}(2016)Newell, Dowdell, and
  Santamore]{Newell2016Surface}
A.~N. Newell, D.~A. Dowdell and D.~H. Santamore, \emph{Journal of Applied
  Physics}, 2016, \textbf{120}, 383--182\relax
\mciteBstWouldAddEndPuncttrue
\mciteSetBstMidEndSepPunct{\mcitedefaultmidpunct}
{\mcitedefaultendpunct}{\mcitedefaultseppunct}\relax
\EndOfBibitem
\bibitem[Hu \emph{et~al.}(2012)Hu, Liu, Xu, and Pan]{hu2012influence}
X.~Hu, G.-Q. Liu, Z.-C. Xu and X.-Y. Pan, \emph{Chinese Physics Letters}, 2012,
  \textbf{29}, 24210--024210\relax
\mciteBstWouldAddEndPuncttrue
\mciteSetBstMidEndSepPunct{\mcitedefaultmidpunct}
{\mcitedefaultendpunct}{\mcitedefaultseppunct}\relax
\EndOfBibitem
\bibitem[Myers \emph{et~al.}(2014)Myers, Das, Dartiailh, Ohno, Awschalom, and
  Bleszynski~Jayich]{Myers2014Probing}
B.~A. Myers, A.~Das, M.~C. Dartiailh, K.~Ohno, D.~D. Awschalom and A.~C.
  Bleszynski~Jayich, \emph{Physical Review Letters}, 2014, \textbf{113},
  027602\relax
\mciteBstWouldAddEndPuncttrue
\mciteSetBstMidEndSepPunct{\mcitedefaultmidpunct}
{\mcitedefaultendpunct}{\mcitedefaultseppunct}\relax
\EndOfBibitem
\bibitem[Romach \emph{et~al.}(2015)Romach, M{\"u}ller, Unden, Rogers, Isoda,
  Itoh, Markham, Stacey, Meijer,
  Pezzagna,\emph{et~al.}]{romach2015spectroscopy}
Y.~Romach, C.~M{\"u}ller, T.~Unden, L.~Rogers, T.~Isoda, K.~Itoh, M.~Markham,
  A.~Stacey, J.~Meijer, S.~Pezzagna \emph{et~al.}, \emph{Physical review
  letters}, 2015, \textbf{114}, 017601\relax
\mciteBstWouldAddEndPuncttrue
\mciteSetBstMidEndSepPunct{\mcitedefaultmidpunct}
{\mcitedefaultendpunct}{\mcitedefaultseppunct}\relax
\EndOfBibitem
\bibitem[Jamonneau \emph{et~al.}(2016)Jamonneau, Lesik, Tetienne, Alvizu,
  Mayer, Dr{\'e}au, Kosen, Roch, Pezzagna,
  Meijer,\emph{et~al.}]{jamonneau2016competition}
P.~Jamonneau, M.~Lesik, J.~Tetienne, I.~Alvizu, L.~Mayer, A.~Dr{\'e}au,
  S.~Kosen, J.-F. Roch, S.~Pezzagna, J.~Meijer \emph{et~al.}, \emph{Physical
  Review B}, 2016, \textbf{93}, 024305\relax
\mciteBstWouldAddEndPuncttrue
\mciteSetBstMidEndSepPunct{\mcitedefaultmidpunct}
{\mcitedefaultendpunct}{\mcitedefaultseppunct}\relax
\EndOfBibitem
\bibitem[Rabeau \emph{et~al.}(2006)Rabeau, Reichart, Tamanyan, Jamieson,
  Prawer, Jelezko, Gaebel, Popa, Domhan, and Wrachtrup]{rabeau2006implantation}
J.~Rabeau, P.~Reichart, G.~Tamanyan, D.~Jamieson, S.~Prawer, F.~Jelezko,
  T.~Gaebel, I.~Popa, M.~Domhan and J.~Wrachtrup, \emph{Applied Physics
  Letters}, 2006, \textbf{88}, 023113\relax
\mciteBstWouldAddEndPuncttrue
\mciteSetBstMidEndSepPunct{\mcitedefaultmidpunct}
{\mcitedefaultendpunct}{\mcitedefaultseppunct}\relax
\EndOfBibitem
\bibitem[Zhou \emph{et~al.}(2017)Zhou, Wang, Zhang, Li, Cai, and
  Gao]{zhou2017self}
Y.~Zhou, J.~Wang, X.~Zhang, K.~Li, J.~Cai and W.~Gao, \emph{Physical Review
  Applied}, 2017, \textbf{8}, 044015\relax
\mciteBstWouldAddEndPuncttrue
\mciteSetBstMidEndSepPunct{\mcitedefaultmidpunct}
{\mcitedefaultendpunct}{\mcitedefaultseppunct}\relax
\EndOfBibitem
\end{mcitethebibliography}

\providecommand*{\mcitethebibliography}{\thebibliography}
\csname @ifundefined\endcsname{endmcitethebibliography}
{\let\endmcitethebibliography\endthebibliography}{}

\end{document}